\documentclass{epl}
\usepackage{epsf}
\usepackage{dcolumn}
\usepackage{bm}

\newcommand\BiCuVO{BiCu$_2$VO$_6$~}

\title{BiCu$_2$VO$_6$: a new narrow-band spin-gap material}
\author{T. Masuda\inst{1}, A. Zheludev\inst{1} \and H. Kageyama\inst{2} \and
A. N. Vasiliev\inst{2,3}\footnote{Permanent address is Low Temperature Physics Department, 
Moscow State University, Moscow 119992, Russia}} 
\institute{
  \inst{1} Condensed Matter Science Division, Oak Ridge National
Laboratory, Oak Ridge, TN 37831-6393, USA \\
  \inst{2} Institute for Solid State Physics, University of
Tokyo,  5-1-5 Kashiwanoha, Kashiwa, Chiba 277-8581, Japan \\
  \inst{3} Department of Low Temperature Physics M.V. Lomonosov
Moscow State University Moscow 119899 Russia 
}


\begin{document}

\maketitle

\begin{abstract}
A new spin-ladder family material BiCu$_2$VO$_6$ is studied by
means of the magnetic susceptibility, heat capacity and neutron
inelastic scattering measurements on powder sample. Singlet ground
state and a finite spin gap are confirmed by thermal-activated
type susceptibility and by distinct peak at 16 meV in spin
excitation. Triple narrow band structure in spin excitation
spectrum, probably due to complex crystal structure, is observed
and the possibility of weakly-interacting spin-cluster system is
discussed.
\end{abstract}

\section{Introduction}

In recent years quantum antiferromagnets with an intrinsically
disordered (``spin liquid'') ground state and energy gap in the
spin excitation spectrum have received a great deal of
attention~\cite{Haldane83a,Dagotto92,Rice93,Hase93a}. Among the
simplest materials of this type are weakly interacting singlet
spin clusters, such as dimers, spin-rings or plaquettes. The
bandwidth of spin excitations from disordered ground state, what
we will call ``magnons'' hereafter, in such materials are
typically smaller or comparable to the gap energy. This feature
make them attractive to both theorists and experimentalists. In
many cases, inter-cluster interactions are sufficiently weak to be
studied perturbatively, yet strong enough to give rise to complex
cooperative phenomena are found in wide-band systems, such as spin
chains and spin ladders. Many prototypical coupled-dimer materials
turned out to be particularly interesting from this point of view.
Among the more extensively studied systems are the alternating
chain compound VODPO~\cite{Garrett97a}, the spin-bilayer material
BaCuSi$_2$O$_6$ ~\cite{Sasago97}, and the
Shastry-Sutherland~\cite{Shastry81} material
SrCu$_2$(BO$_3$)$_2$~\cite{Kageyama99}. Neutron scattering
experiments on Cu(NO$_3$)$_2$2.5D$_2$O~\cite{Xu00} revealed not
only dispersive single-dimer modes, but also a highly structured
continuum of multi-dimer excitations. Another interesting case is
that of 
(Tl,K)CuCl$_3$~\cite{Oosawa99a, Cavadini02, Oosawa02a,Oosawa02b,Oosawa03a}, where
antiferromagnetic long-range order has been induced by a strong
external magnetic field, pressure, or impurity doping.

In our search for new experimental realizations of spin-ladder and
related models, we came across a new interesting family of
low-dimensional materials with the general formula
BiCu$_2$$A$O$_6$ ($A$ = V~\cite{Radosavljevic98},
P~\cite{Abraham94}, and As~\cite{Radosavljevic99}). As will be
shown in the present paper, at least one of these compounds,
namely BiCu$_2$VO$_6$, indeed has a non-magnetic singlet ground
state and a gap in the spin excitation spectrum.

\begin{figure}
\begin{center}
\epsfxsize=12cm \epsfclipon \epsffile{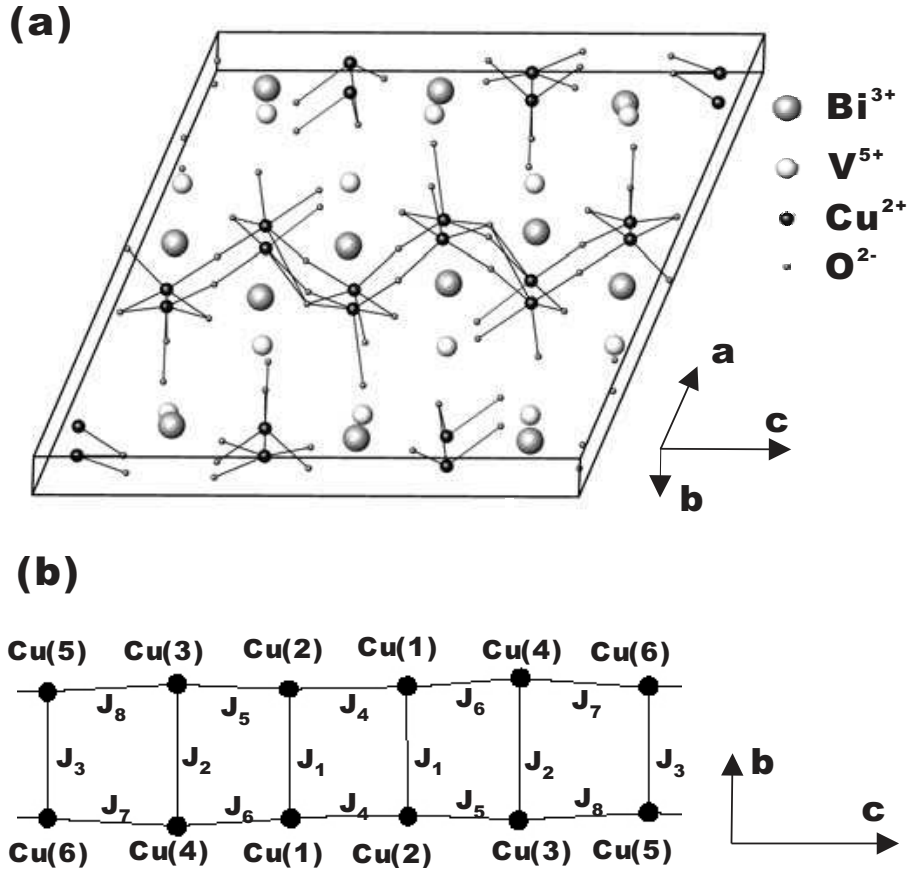}
\end{center}
\caption{ (a) Crystal structure of BiCu$_2$VO$_6$. $S$ = 1/2
carrying Cu$^{2+}$ ions form zig-zag ladder along the
crystallographic $c$ axis, which is separated by non-magnetic
Bi$^{3+}$ and V$^{5+}$ ions. (b) Framework of Cu$^{2+}$ ions
projected onto $b$ - $c$ plane. Six inequivalent Cu$^{2+}$ ions
suggests eight different exchange interaction in distorted ladder
structure.} \label{fig1}
\end{figure}

The crystal structure of \BiCuVO is of relatively low symmetry,
monoclinic, space group $P2_1/n$~\cite{Radosavljevic98}, and is
schematically visualized in Fig.~\ref{fig1} (a). Lattice parameters are 
$a$ = 13.49 \AA, $b$ = 7.822 \AA, $c$ = 15.79 \AA, and $\beta$ = 
113.113 $^{\circ}$. The most
prominent feature are zig-zag ladder structures formed by the
$S=1/2$-carrying Cu$^{2+}$, assumed to be responsible for the
magnetic properties. The ladders run along the crystallographic
$c$ - axis and are separated by  non-magnetic V$^{5+}$ and
Bi$^{3+}$ ions. The rung of the ladders are roughly parallel to
the crystallographic $b$ - direction. As shown in Fig.~\ref{fig1}
(b), the ladders  actually incorporate six {\it inequivalent}
Cu-sites, which results in eight {\it inequivalent} Cu-O-Cu bond. 
Given the location of O$^{2-}$ ions that are likely to
play a key role in superexchange interactions in this material,
one can hypothesize a spin Hamiltonian with up to eight unequal
exchange parameters (Fig.~\ref{fig1} (b)). Even though
crystallographic considerations point towards a
quasi-one-dimensional distorted ladder-type system, we find that
\BiCuVO\ possesses all the features of a narrow-band ``cluster''
model, possibly composed of weakly interacting dimers.

\section{Experimental Procedure and Results}
\begin{figure}
\begin{center}
\epsfxsize=12cm \epsfclipon \epsffile{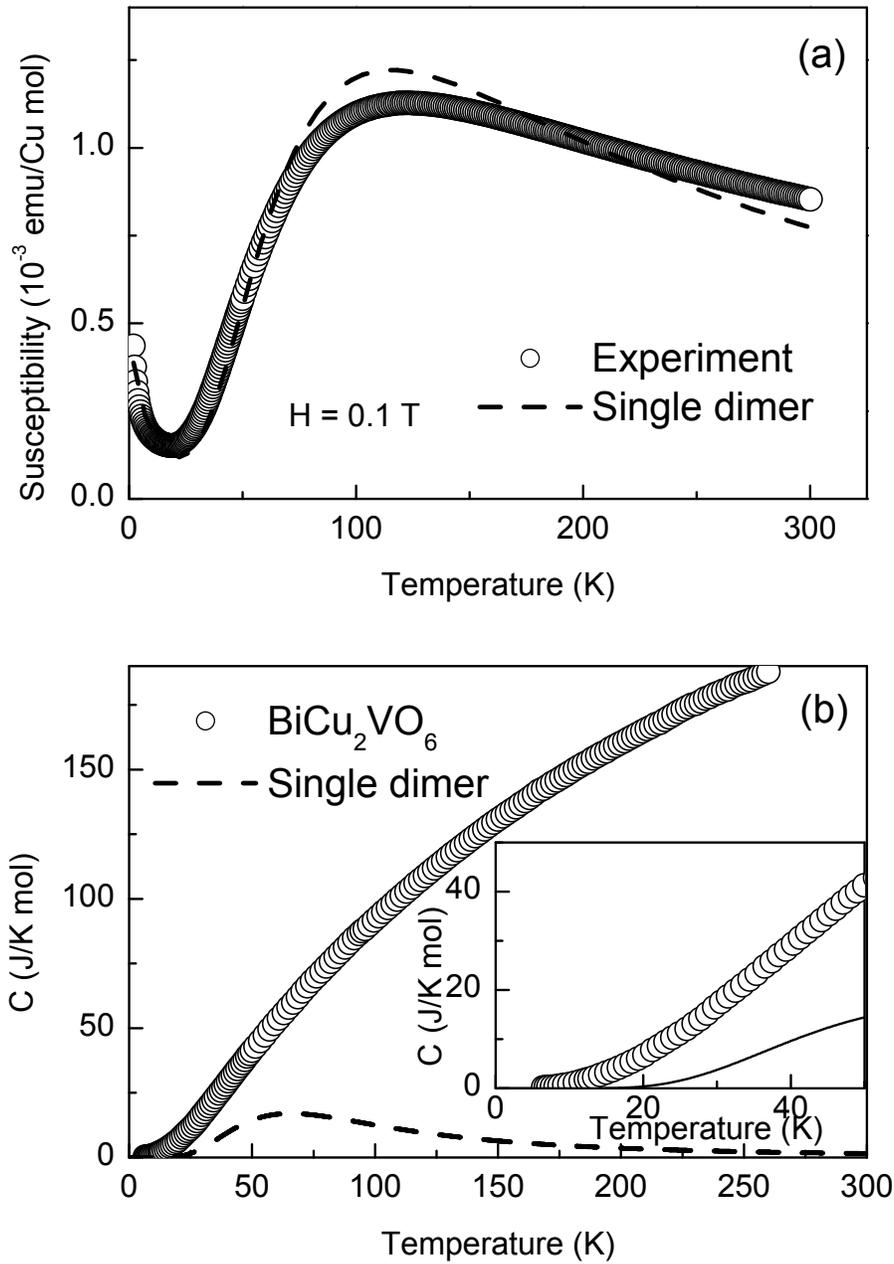}
\end{center}
\caption{(a) Magnetic susceptibility (open circle) and fit by
spin-dimer model (dashed line). The thermal-activated feature
suggests the singlet ground state and a finite spin gap. Applied
field is 0.1 T and a temperature range is 2 - 300 K. (b) Heat
capacity (open circle) and spin-heat capacity predicted by
spin-dimer model (dashed line). No signature of phase transition
is observed. Applied field is 0 T and a temperature range is 6.4 -
260 K} \label{fig2}
\end{figure}

The non-magnetic nature of the ground state and the energy gap in
\BiCuVO were detected in bulk susceptibility measurements. These
data were collected using a Quantum Design SQUID magnetometer on a 
powder sample in a 1000 Oe field, and are shown in Fig.~\ref{fig2}
(a). Apart from the low-temperature upturn that is most likely due
to impurities, the susceptibility curve has a  pronounced
thermal-activated character. As a point of reference, it can be
approximated by a $S=1/2$ dimer susceptibility curve, with an
additional Curie contribution at low temperatures due to
impurities. The result of such fit shown in a solid line
in~\ref{fig2} (a). In plotting this curve we used the value
$g=2.03$ for the gyromagnetic ratio of Cu$^{2+}$, as determined by
room-temperature ESR measurements. The fitted a singlet-triplet
dimer splitting is $\Delta 16$~meV. The amount of impurity was 
estimated to be only 0.8\% of Cu$^{2+}$ ions. While this simplest 
local-singlet models is probably an inappropriate description of
\BiCuVO, the extracted gap energy sets the characteristic
magnitude of spin interatcions.

Due to the large magnetic energy scales involved, the
magnetic contribution to heat capacity is too weak to be isolated from
the phonon part. This fact is borne out in Figure ~\ref{fig2} (b)
that shows the $C(T)$ curve measured in a "Termis" calorimeter using 
the quasiadiabatic method on a polycrystalline sample a 
polycrystal sample (circles), in comparison with that calculated
for the dimer model described above (solid line). One useful piece
of information that is contained in the specific heat data, is the
absence of any signatures of phase transitions over a temperature
range 6.4 - 260 K. This supports the notion that the non-magnetic
ground state is an intrinsic property of the spin network as it is
at room temperature, and not a result of some structural
transition.

More insight into the nature of the ground state and magnetic gap
excitations was obtained in inelastic neutron scattering
experiments on \BiCuVO. The data were collected on a 14.5g powder
sample using the HB-1 3-axis spectrometer installed at the High
Flux Isotope Reactor at Oak Ridge national Laboratory. Neutrons of
a fixed final energy $E_f=13.5$~meV were used in combination with
a graphite filter positioned after the sample. Soller collimators
defined the neutron beam divergencies as $48' - 60' - 60' - 120'$.
Sample environment was a standard closed-cycle refrigerator.

\begin{figure}
\begin{center}
\epsfxsize=12cm \epsfclipon \epsffile{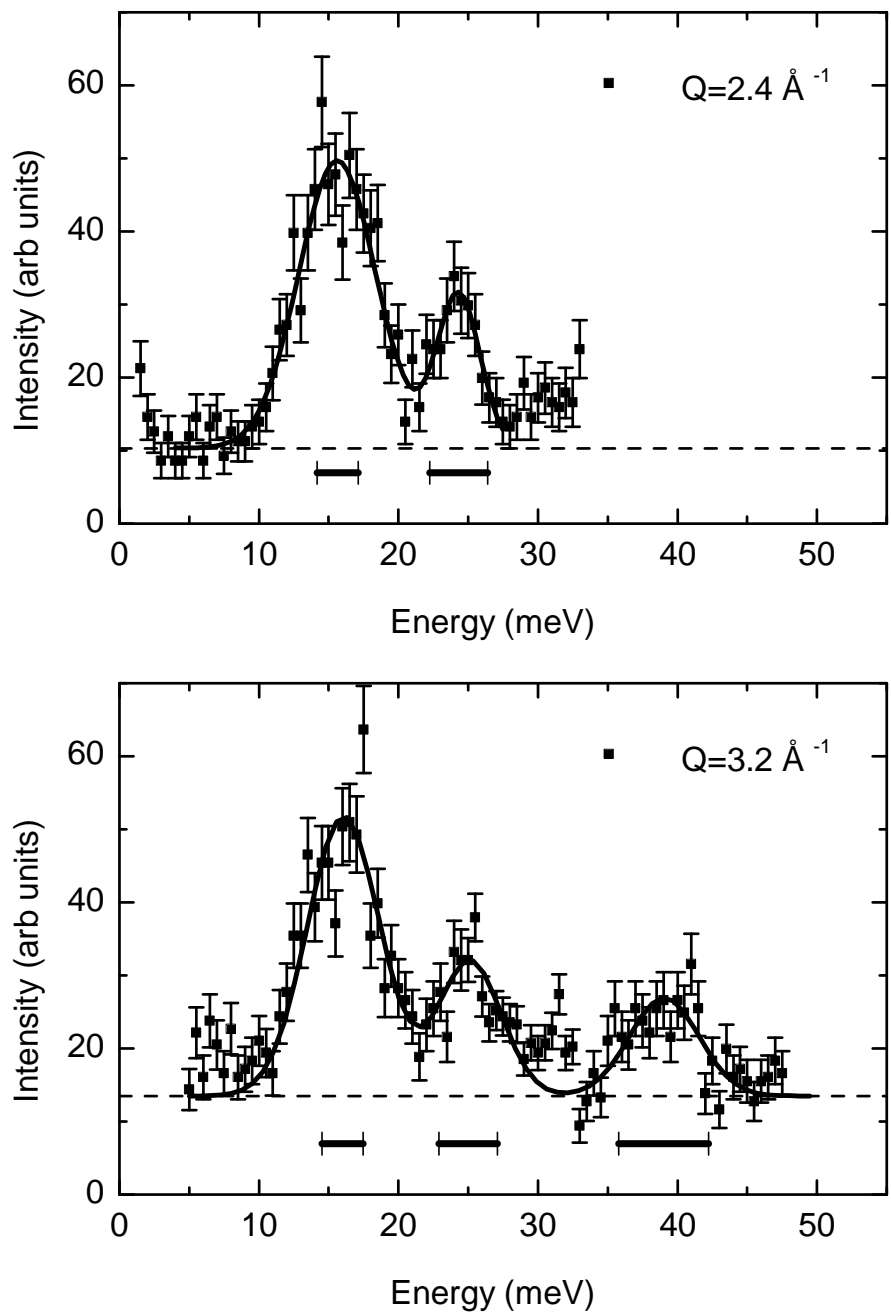}
\end{center}
\caption{Constant Q scan at 2.4 \AA $^{-1}$ and
 3.2 \AA $^{-1}$. Solid line is fit by multi Gaussian,
 dotted line is background from the fit, bars are resolution in energy.
 Distinct peaks are observed at 16, 25, and 39 meV. The peak at 16 meV
 has intrinsic finite width while those at 25 and 39 meV are
 in resolution limit. }
\label{fig3}
\end{figure}

\begin{figure}
\begin{center}
\epsfxsize=12cm \epsfclipon \epsffile{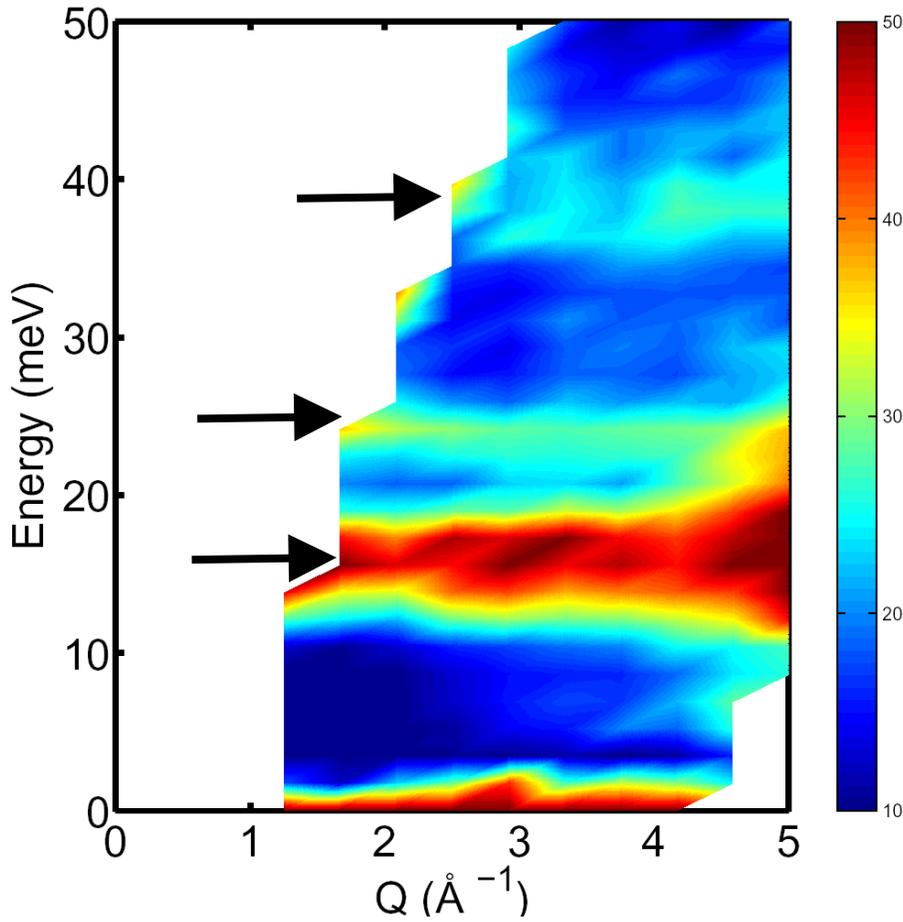}
\end{center}
\caption{Energy vs Q false color map. Triple narrow bands are
indicated by arrows. The absence of intensity below lowest energy
band at 16 meV means the existence of bound state there and
intrinsic spin gap in this material. Triple band structure is
discussed in the text. } \label{fig4}
\end{figure}

Typical constant-$Q$ scans collected at $T=10$~K are shown in
Fig.~\ref{fig3}. A number of such scans were combined to produce
the false-color plot shown in Fig.~\ref{fig4}. Inelastic powder
data are notoriously difficult to interpret due to the spherical
averaging that is an intrinsic feature of powder scattering.  In
our case of \BiCuVO the situation is even more complicated than
usual, largely due to the energy scales involved. The
powder-averaged dynamic structure factor is most informative at
small momentum transfers, where the averaging occurs across only a
few Brillouin zones, and where the signal often has a lot of
``structure''. A good example of such behavior was witnessed in
recent powder experiments on the Haldane-gap antiferromagnet
 PbNi$_2$V$_2$O$_8$\cite{Uchiyama99,Zheludev00a,Zheludev01a}. In the present study however, momentum
transfers below 1~\AA$^{-1}$ are unreachable in the interesting
energy transfer range (15 to 40 meV) due to kinematic
constraints~\cite{kinematics}. At higher wave vectors, any
characteristic features of the cross section are progressively
``smeared'' by spherical averaging over several Brillouin zones.
Magnetic scattering then becomes similar to the magnon density of
states function. Moreover, the background, largely due to
inelastic incoherent scattering by phonons that involve the
V-nuclei, becomes progressivly stronger at higher momentum
transfers, while magnetic scattering is suppressed due to the
magnetic form factor.

Given these intrinsic limitations of the present powder experiment
and the complexity of the crystal structure and geometry of
magnetic interactions, a quantitative analysis of the data,
similar to the one described in
Refs.~\cite{Uchiyama99,Zheludev00a,Zheludev01a} does not appear
feasible. The data obtained in our neutron experiments can
nevertheless provide some qualitative clues to the nature of the
singlet ground state and energy gap in \BiCuVO. The most prominent
feature of the inelastic scattering, clearly visible in
Fig.~\ref{fig4}, are three bands of intensity at energy transfers
16 meV, 25 meV, and 39 meV, respectively. The central energies
were determined in Gaussian fits as those shown in solid lines in
Fig.~\ref{fig3}. Overall, in the admittedly narrow $q$-range
sampled, there is very little wave vector dependence to
scattering. As shown in Fig.~\ref{fig3}, only the lower peak is
somewhat broader than the experimental energy resolution
(intrinsic width 3.2~meV FWHM), and the two higher-energy features
are resolution-limited. The 16~meV peak is clearly to be
associated with the energy gap in the system.

\section{Discussion}

There are at least two alternative explanations for the observed
intensity bands. One possibility is that the scattering is due to
a single dominant magnon branch with a rather wide dispersion
bandwidth between the gap energy of $\Delta\approx 16$~meV and a
zone-boundary energy of $39$~meV. The three peaks are then
associated with Van Hove singularities in the magnon density of
states. The location and intensity of the peaks primarily depends
on the details of the magnon dispersion relation. Such is the case
for powder experiments on several materials including
PbNi$_2$V$_2$O$_8$.\cite{Uchiyama99,Zheludev00a,Zheludev01a,DiTusa94a,DiTusa94b}
A different explanation attributes the three observed peaks to
three distinct branches of narrow-band singlet-triplet excitations
in one or more local spin clusters, such as dimers. For \BiCuVO
this explanation seems to be the more likely one. Indeed the
measured powder spectrum (Fig.~\ref{fig3}) has many similarities
with that seen in the well-characterized isolated-dimer
material~\cite{Tennant97a} (See Fig.~4 in Ref.~\cite{Tennant97a}
). The observed $q$-dependence is a slowly varying function that
may represent smooth oscillations of the single-cluster structure
factor, rather than sharp features expected from powder-averaging
of magnon cross sections with a steep dispersion
\cite{Uchiyama99,Zheludev00a,Zheludev01a,DiTusa94a,DiTusa94b}. In
the local-cluster interpretation the intrinsic width of the 16~meV
peak in \BiCuVO is attributed to weak interactions between the
clusters, the magnon bandwidth being directly related to the
apparent energy width of the intensity band in the powder data.
The symmetry of \BiCuVO is low enough to allow up to three
different dimer types, based on the 6 inequivalent Cu$^{2+}$
positions. Other possibilities may include 4- and even 6-spin
clusters. Lacking a reliable model of magnetic interactions in
\BiCuVO, a further exploration of this avenue will have to be
postponed until single crystal samples become available.

\section{Conclusion}
There are two lessons to be learned from our experiments on
\BiCuVO powder samples. First, we have shown that at least one
member of the BiCu$_2$$A$O$_6$ family is a singlet ground state
material, and identified the underlying ladder-type spin
arrangement. Second, it became obvious that the main obstacle to
fully understanding the physics of \BiCuVO is in its distorted and
complicated crystal structure. Other members of the family, for
example BiCu$_2$$P$O$_6$ may be more valuable as model quantum
magnets, since their structure features undistorted spin ladders.

\section{Acknowledgements}

We thank Dr. E. Popova for assistance in specific heat 
measurements and Dr. Y. Petrusevich and Dr. Y. Koksharov 
for taking ESR spectra at room temperature. 
Work at ORNL  was carried out under Contracts No.
DE-AC05-00OR22725, US Department of Energy.
Work at ISSP was supported by 
the MECSST (Japan) through Grant-in-Aid for Scientific
Reserach No. 40302640.
A.N.V. acknowledge support through RFBR Grant 
03-02-16108 and NWO Grant 008-012-047.

\begin{thebibliography}{99}
\bibitem{kinematics}
Kinematics constraints, $\delta E = \frac{\hbar}{2m}(2{\mbox{\boldmath
$k_i$}}\cdot {\mbox{\boldmath $q$}}+q^2)$, restricts the maximum energy
transfer small at small $q$. Here $\delta E$ is
energy transfer, {\boldmath $k_i$} is wave vector of incident
neutron, {\mbox{\boldmath $q$}} is scattering vector, and $m$ is
the mass of neutron.
\end{thebibliography}
\end{document}